**Probing subtle fluorescence dynamics in cellular proteins by streak camera based Fluorescence Lifetime Imaging Microscopy**

R.V. Krishnan*, Eva Biener, Jian-Hua Zhang, Robert Heckel and Brian Herman

Department of Cellular and Structural Biology,
University of Texas Health Science Center,
7703 Floyd Curl Drive, San Antonio, Tx 78229, USA.

* Corresponding author : krsna@uthscsa.edu

**Abstract**

We report the cell biological applications of a recently developed multiphoton fluorescence lifetime imaging microscopy system using a streak camera (StreakFLIM). The system was calibrated with standard fluorophore specimens and was shown to have high accuracy and reproducibility. We demonstrate the applicability of this instrument in living cells for measuring the effects of protein targeting and point mutations in the protein sequence – which are not obtainable in conventional intensity-based fluorescence microscopy methods. We discuss the relevance of such time-resolved information in quantitative energy transfer microscopy and in measurement of the parameters characterizing intracellular physiology.

Wide field fluorescence microscopy methods measure fluorescence emission in two-dimensions, on a pixel-by-pixel basis. Optical sectioning methods, such as confocal laser scanning microscopy, make it possible to scan the (x,y) images along the z-axis thereby allowing three-dimensional reconstruction of the specimen.[1,2] Fluorescence lifetime imaging microscopy (FLIM) extends the scope of these intensity-based fluorescence microscopy methods by revealing a vital parameter, fluorescence lifetime ($\tau$), providing an extra dimension in live cell imaging. Unlike intensity-based measurements the fluorescence lifetime is independent of spectral similarity and concentration of multiple fluorophore labels employed.[3,4,5]

In this letter we describe a recently developed multiphoton FLIM system which uses a streak camera and report some novel applications of this system in cell biology. This FLIM system combines high spatial and temporal resolution through the coupling of multiphoton excitation and streakscope detection.[6] We demonstrate the applicability of this instrument in living cells for measuring the effects of organelle targeting and single amino acid changes in one of the fluorescent proteins – enhanced cyan fluorescent protein (ECFP)- which are not obtainable in conventional intensity-based fluorescence microscopy methods. We have chosen to examine ECFP because of its wide use in cellular imaging.[7] Besides being an endogenous probe for protein tracking in living cells, ECFP is often used as a donor for fluorescence resonance energy transfer (FRET) measurements where it is paired with either EYFP or DsRed as the acceptor.[8] Since FRET microscopy is being employed extensively for measuring protein-protein interactions, information about the fluorescence dynamics of the donor/acceptor will not only improve the accuracy of FRET detection, but will also provide a better way for quantification of the parameters characterizing intracellular physiology. With this motivation, we addressed three questions pertaining to the fluorescence dynamics of this protein : (i) Does the construction of fusion protein affect the measured lifetime of ECFP? (ii) Is the lifetime of ECFP different when targeted to specific organelles inside the cell? and (iii) Can lifetime measurements detect the changes in fluorescence decay due to single amino acid changes in ECFP protein sequence? Two model cell lines, baby hamster kidney (BHK-21) and human embryonic kidney (HEK 293T), were used in these studies. Cells were grown on glass coverslips in dishes and transfected using FuGENE 6 as per the manufacturer's (Roche Applied Science, Indiana) guidelines, with DNA constructs encoding the following fluorescent proteins : cytosolic expressed enhanced CFP (ECFP), cytosolic expressed fusion protein Bax-ECFP, membrane targeted ECFP (mem-ECFP), mitochondrially targeted ECFP (mito-ECFP) and cytosolic expressed ECFP with a single amino acid change (A207K) (mut-ECFP) that is known to reduce dimerization of fluorescent proteins.[9]

Figure 1 shows the configuration of the streak FLIM imaging system. The Ti:Sapphire laser (Model Mira 900, Coherent Inc.) provides mode-locked, ultrafast femtosecond pulses with a fundamental frequency of 76 MHz. A pulsepicker (Model 9200, Coherent Inc.) is used to produce variable repetition rates ranging from 146 kHz to 4.75 MHz. In the FLIM path, the output pulses of the pulsepicker have a repetition rate of 500 kHz and power typically around 8 mW at the entrance of the FLIM optics. An Olympus IX-70 inverted microscope with a specially designed 63X (1.2 N.A, IR) water immersion objective was used. The fluorescence emission from the microscope is collected by G2 and transmitted through D1 for detection by the streakscope. The streakscope consists of a photocathode surface, a microchannel plate photomultiplier tube (MCP-PMT) to amplify photoelectrons coming off the photocathode and a phosphor screen to detect the photoelectrons. A brief description of streak imaging is given in Figure 1 and more details can be found in Reference 6. StreakFLIM measurements were carried out on transfected cells grown on glass coverslips which were fixed and mounted on a standard microscope slide. To obtain correlative results using a different FLIM method, parallel measurements were made using a commercially available time-correlated single photon counting (TCSPC) FLIM module (SPC–730, Becker & Hickl Inc.) coupled with the laser scanning confocal microscope (Zeiss LSM510-NLO) equipped with a Ti:Sapphire laser for multiphoton excitation.[10] The standardization of both these FLIM systems was done using a standard fluorophore solution (Rhodamine 6G in ethanol) and the lifetime values obtained with either system agree with previously reported results.[3]

Crucial requirements for live cell FLIM imaging are rapid data acquisition, minimal photobleaching, high accuracy and reproducibility in measurement. Table 1 compares the typical imaging parameters used in StreakFLIM and TCSPC systems. In spite of having lower excitation probability (500 kHz vs. 76 MHz), a larger image size to scan and a smaller numerical aperture of the objective lens (which governs both excitation and detection probabilities) it was found that the StreakFLIM system could acquire data at least three times faster than TCSPC system. It is possible to decrease the data acquisition time by another two-fold in the present StreakFLIM conditions by decreasing the size of the acquired image. Considering the stochastic nature of fluorescence emission (typically 1 fluorescence photon is emitted for every $10^3$ excitation pulses), it was found that longer data integration (without photobleaching) improves the signal-to-noise ratio and enhances the optical content from the specimen. All the measurements were done in optimal imaging conditions and the changes in lifetime reported in this letter were above the instrument sensitivity limit (0.08-0.1 ns).[6] A similar sensitivity and lifetime resolution could be expected from a more established time-domain FLIM system based on multiple time-gated detection. Gerritsen and coworkers have demonstrated earlier that such a lifetime module enables the recording of fluorescence lifetimes by capturing the intensity decay in four or eight time gates at high count rates and with a high efficiency.[4] Using a fast photomultiplier tube operated close to its maximum count rate, these authors were able to record a FLIM image (256 x 256 pixels) of a beating myocyte in ~ 1s. In a nutshell, the final lifetime resolution in any FLIM system is always dominated by photon statistics and time response of the detectors employed.

Figure 2 shows the representative lifetime images of ECFP expressed in BHK and HEK cells. First of all, the mean lifetime value of cytosolic expressed ECFP is 3.17 ± 0.26 ns in both BHK and HEK cell lines (2a & 2g) demonstrating the intrinsic nature of measured ECFP lifetime as a quantitative parameter. Identical mean lifetime values (3.20 ± 0.28 ns) were obtained from these specimens independently by the TCSPC system (Figure 2c). Figure 2 (b-f) also shows the lifetime images and histograms of cells expressing ECFP either fused to a protein (Bax-ECFP) or targeted to specific organelles (mem-ECFP and mito-ECFP). As can be seen from Figure 2, there is a considerable reduction in ECFP lifetime in these cells as compared to the free cytosolic ECFP (ECFP: 3.17 ± 0.26 ns; Bax-ECFP: 2.80 ± 0.27 ns; mem-ECFP: 2.90 ± 0.09 ns; mito-ECFP: 2.91

± 0.07 ns). These statistically significant changes demonstrate the sensitivity of lifetime measurements in detecting the effects of expressing ECFP as a fusion protein or adding a targeting sequence to the N-terminus of the CFP. Intensity-based fluorescence microscopy methods are insensitive to these subtle fluorescence dynamic changes and it is therefore not possible to obtain this information from such methods. We did find that there was no significant difference in lifetimes between different organelles (e.g. mito-ECFP vs. mem-ECFP). We found similar changes in lifetime in many other fusion proteins and organelle targeted (nuclear ECFP etc.,) other than shown in Figure 2 (data not shown). Our observations are therefore generic in nature and demonstrate an important finding that lifetime measurements can indeed report subtle fluorescence dynamics arising from even minor modifications in the protein sequence. It should be noted that the observed changes in lifetimes do not arise from intensity artifacts but are solely from changes in molecular fluorescence.

Another important molecular change that is relevant in biological sciences is mutation in the amino acid sequence of a protein that can change the overall structure/property relationship of the protein dramatically. To investigate the possibility of measuring the effects of such single amino-acid changes, we used site-directed mutagenesis in the ECFP by changing Alanine to Lysine (A207K). This mutation is thought to decrease the tendency of the ECFP to dimerize although there is no documentation about the effect this mutation can have on the fluorescence of ECFP. Figure 2(g)-(i) shows the lifetime images and histogram of HEK cells expressing wildtype ECFP and mutant ECFP (A207K). An almost identical expression level (fluorescence emission intensity) was observed in both the cases thereby making it impossible to discern any difference from intensity images (data not shown). However, there was a clear reduction in lifetime in the A207K mutant ECFP ($\tau = 2.89 \pm 0.22$ ns) as compared to the ECFP ($\tau = 3.17 \pm 0.26$ ns), demonstrating the dramatic sensitivity of our FLIM systems to subtle molecular changes in fluorescent proteins. It is not clear how a single amino acid change can affect the fluorescence lifetime so drastically. We speculate that a single amino acid change introduces some charge interactions or some other interactions such as steric clearance issues that can affect the local environment of the fluorophore which, in turn, results in a difference in fluorescence lifetime without affecting the wavelength of the emission. This is probably the reason why intensity images do not show a discernible difference between ECFP and mutant (A207K) ECFP. A detailed structural analysis may throw more light on the effects of amino acid changes in modifying fluorescence decay scheme. However these studies are not within the scope of this letter.

Our purpose in this report is to highlight that the measurement of fluorescence dynamics of cellular proteins with high spatial/temporal resolution (unlike spectroscopic ensemble measurements) is now possible. Furthermore, we envisage that such measurement capability can yield useful guidelines in designing proper controls for quantitative live cell imaging such as FRET microscopy and in situ pH imaging.

We gratefully acknowledge Dr. Brad Rothberg for his generous gift of mem-ECFP construct used in this paper. This work was supported by NIH grant AG20210.

**Table Legend**

**Table 1 : A comparison of imaging parameters for estimating the relative efficiency of StreakFLIM and TCSPC FLIM systems using multiphoton excitation (840nm)**

| Imaging Parameters | StreakFLIM | TCSPC |
|---|---|---|
| Excitation Pulse repetition rate | 500 kHz | 76 MHz |
| Average power at the sample | 8 mW | 25 mW |
| Objective lens | 60 X, 1.2 NA | 63X, 1.4 N.A |
| Image size | 658 x 494 | 256 x 256 |
| Pixel dwell time | 3 µs/pixel | 3 µs/pixel |
| Average data acquisition time | ~ 22 sec | ~ 60 sec |

**Figure Captions**

**Figure 1: Schematic of multiphoton StreakFLIM system.**

G1: Horizontal Galvo Mirror; G2 : Vertical Galvo Mirror ;  L1: Focusing relay lens; C: Coupling lens for sideport optics; I : Imaging lens; O : Microscope Objective lens;  D1 : Dichroic(Short pass filter/cutoff 750 nm);  PC :Photocathode;    As the beam scanner G1 scans a single excitation spot along a single line across the x-axis (width of (x,t) line ~ 658 points), the streakscope collects the time decay of fluorescence emission from every point along this line.   Individual electrons arriving at the sweep electrode at different times are deflected at different angles in the vertical direction and therefore, one obtains a streak image with space and time as the ordinate and abscissa respectively.   When synchronous y-scanning (G2) is carried out on the region of interest, the above streak imaging process gives a complete stack of (x,t) streak images.  This stack contains the complete information of optical intensity as well as the spatial and temporal information from the optical image.   Numerical processing of all these streak images (i.e. the exponential decay profiles at every pixel of the raw streak image) gives the final FLIM image.  The insets show the lifetime image and histogram for the standard fluorophore solution of Rhodamine 6G in ethanol.  Also shown is the fluorescence decay for the same specimen measured in time-correlated single photon counting (TCSPC) system.   The literature reported lifetime value for this specimen is 3.0 ns.

**Figure 2: Representative Fluorescence lifetime images and histograms**

**(a) & (b)** : BHK cells expressing cytosolic free ECFP and the fusion protein Bax-ECFP; **(c)** Lifetime histogram comparing cytosolic ECFP and Bax-ECFP measured in StreakFLIM; also shown is the lifetime histogram of cytosolic ECFP measured with TCSPC FLIM; **(d) & (e)** Lifetime images of cells expressing organelle (mitochondrial and membrane) targeted ECFP; **(f)** Lifetime histograms of free and organelle targeted ECFP; also shown is the lifetime histogram of cytosolic ECFP in BHK and HEK cells; **(g) & (h)** : Lifetime images of HEK cell expressing ECFP and A207K mutant ECFP **(i)** : Lifetime histograms of ECFP and mutant ECFP.

**References**

[1] J.B. Pawley, (ed.) *Handbook of Biological Confocal microscopy, Second Edition* (Plenum, NewYork,1989).

[2] A. Periasamy, (ed.) *Methods in Cellular Imaging*, (Oxford University Press, NewYork,2001).

[3] A. Periasamy, P. Wodnicki, X.F.Wang, S. Kwon, G.W. Gordon and B. Herman, Rev. Sci. Instrum. **67**, 3722 (1996).

[4] H.C. Gerritsen, M.A.H. Asselbergs, A.V. Agronskaia and W.G.J.A.H.M. Van Sark, J.Microsc. **206**, 218 (2002).

[5] P.J. Verveer, F. S. Wouters, A.R.Reynolds and P.I.H.Bastiaens, Science **290,** 1567 (2000).

[6] R.V. Krishnan, H.Saitoh, H.Terada, V.E. Centonze and B.Herman, Rev.Sci.Instrum. **74**, 2714 (2003).

[7] R.Y. Tsien, , *Annu.Rev.Biochem,* **67**, 509 (1998) and references therein;

[8] B.W. Hicks, (ed.) *Green Fluorescent Protein*,(Humana,Totowa,NJ ,2002)

[9] D.A. Zacharias, J.D.Violin, A.C.Newton and R.Y. Tsien, Science **296**, 913 (2002).

[10] W.Becker, A.Bergmann, C.Biskup, L.Kelbauskas, T.Zimmer, N.Klocker, K.Bendorf, Proc .SPIE. **4963,** 175 (2003).

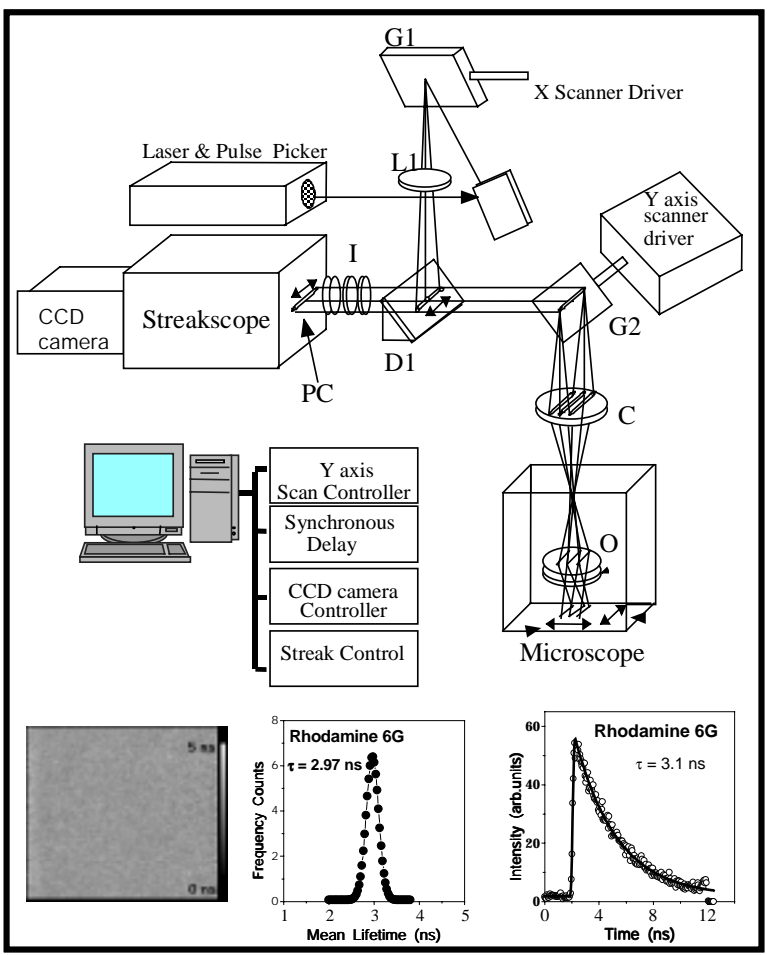

**Figure 1**

**Krishnan et al.**

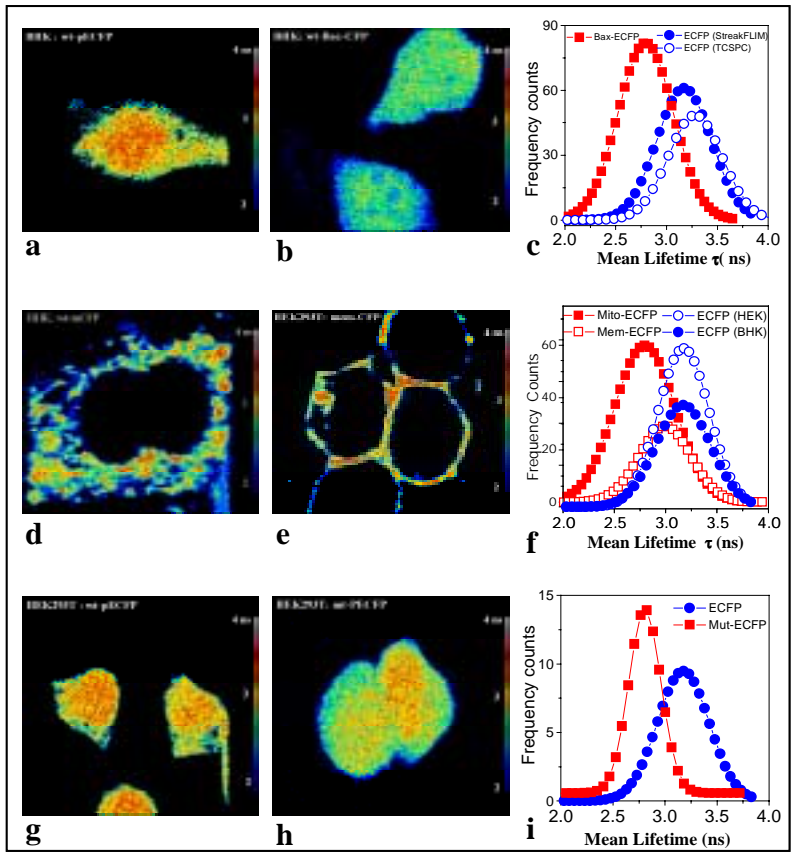

**Figure 2**

**Krishnan et al.**